\documentclass[aps,prb,twocolumn,showpacs,showkeys]{revtex4}
\usepackage{amsmath,bm}
\usepackage{mathrsfs}
\usepackage{amsfonts}
\usepackage{graphicx}
\usepackage{color}
\usepackage{dcolumn}
\newcommand{\ksa}{{\mathbf k,a\sigma}}
\newcommand{\bra}[1]{\left\langle #1 \right| }
\newcommand{\ket}[1]{\left| #1 \right\rangle }

\begin{document}

\title{Many-body effects in magnetic inelastic electron tunneling spectroscopy}
\author{Richard Koryt\'ar}
\affiliation{
Institut f\"ur Nanotechnologie,
Karlsruher Institut f\"ur Technologie, Hermann-von-Helmholtzplatz 1,
D-76344 Eggenstein-Leopoldshafen, Germany
}
\affiliation{Centro de investigaci\'on en nanociencia y nanotecnolog\'{\i}a
(CSIC - ICN), Campus de la UAB, E-08193 Bellaterra, Spain}
\author{Nicol\'as Lorente}
\affiliation{Centro de investigaci\'on en nanociencia y nanotecnolog\'{\i}a
(CSIC - ICN), Campus de la UAB, E-08193 Bellaterra, Spain}
\author{Jean-Pierre Gauyacq}
\affiliation{{
Institut des Sciences Mol\'eculaires d'Orsay, ISMO, Unit\'e mixte
CNRS-Universit\'e Paris-Sud,UMR  8214, B\^atiment 351, Universit\'e
Paris-Sud, 91405 Orsay CEDEX, France}} 
\date{\today}

\begin{abstract}
Magnetic inelastic electron tunneling spectroscopy (IETS) shows sharp
increases in conductance when a new conductance channel associated to
a change in magnetic structure is open. Typically, the magnetic moment carried 
by an adsorbate can be changed by collision with a tunneling electron;
in this process the spin of the electron can flip or not. A previous
one-electron theory [Phys. Rev. Lett. {\bf 103}, 176601 (2009)]
successfully explained both the conductance thresholds and the magnitude
of the conductance variation.
 The elastic spin flip of conduction electrons by a magnetic impurity
leads to the well known Kondo effect.  In the present work, we compare
the theoretical predictions for inelastic magnetic tunneling obtained
with a one-electron approach and with a many-body theory  including
Kondo-like phenomena. We apply our theories to a singlet-triplet
transition  model system that contains most of the characteristics
revealed in magnetic IETS.   We use two self-consistent treatments 
(non-crossing approximation and self-consistent ladder approximation). We 
 show that, although the one-electron limit is properly recovered, 
new intrinsic many-body features appear.  In particular,
sharp peaks appear close to the inelastic thresholds; these are not
localized exactly at thresholds and could influence the determination
of magnetic structures from IETS experiments.Analysis of the evolution with temperature reveals that these 
many-body features involve an energy scale different from that of the usual Kondo peaks.
Indeed, the many-body features perdure at temperatures much larger
than the one given by the Kondo energy scale of the system.

\end{abstract}
\pacs{68.37.Ef,72.15.Qm,72.10.Fk,73.20.Hb} 
\keywords{IETS, spin-flip, Kondo effect, singlet-triplet Kondo}
\maketitle

\section{Introduction}

Magnetic inelastic electron tunneling spectroscopy (magnetic IETS)
detects magnetic excitations on a surface by measuring the changes of
conductance of a scanning tunneling micrsocope (STM) junction when an
excitation is produced~\cite{HeinrichScience2004}. As in general IETS,
{the possibility to excite the surface leads to an increase of the
number of  possible final channels for the tunneling  electron when
the junction bias matches an excitation energy threshold, consequently
leading to an abrupt increase of the junction conductance}.
 This high sensitivity
  {on} the small magnetic energy scale has permitted
{Hirjibehedin and coworkers}~\cite{HeinrichScience2007} to measure the
magnetic anisotropy energy (MAE) of single atomic adsorbates as well as
the magnetic coupling among them~\cite{HeinrichScience2006}. More recent
experiments {revealed}  the change of sign of MAE when a phthalocyanine
molecule is adsorbed on a surface~\cite{TsukaharaPRL2009} and {the
existence of an } exchange coupling between magnetic molecules in a
multilayer setup~\cite{ChenPRL2008}.  This technique has open a venue
to the study of magnetic phenomena {at surfaces} on the single atom {or}
molecule scale.

Theories have been developed to rationalize the steps
in conductance found in magnetic IETS with a good degree of
success~\cite{Fransson,FernandezPRL2009,PerssonPRL2009,Lorente:PRL2009}.
The large changes in conductance observed in magnetic IETS were
explained by the large parent coefficients {of the initial and final
adsorbate states in the tunneling state ~\cite{Lorente:PRL2009}. During
tunneling the spins of the tunneling electron and of the magnetic
adsorbate couple together into a total spin $S_T$ that characterizes
the spin symmetry of the tunneling process.  The excitation process
is then pictured as the coupling/decoupling of the two spins. Since
the parentage coefficients of the $S_T$ states in the initial and final
magnetic states of the adsorbate can be large (these are Clebsch-Gordan
coefficients associated to MAE structure coefficients), the probability
of forming various excited final states can be very large. In this way,
the flux of incident electrons is shared among the possible final
channels that are energetically possible; the branching ratios among channels are
simply  governed by structure coefficients.  This theory also showed
that the magnetic excitation of an adsorbate may or may not imply 
spin-flip of the tunneling electron~\cite{Lorente:PRL2009}.}

The {Kondo effect is associated with a phenomenon bearing many links with
magnetic IETS: the fluctuations induced by the spin-flip of an electron
during its collision with a magnetic impurity at constant energy.} Kondo physics
and inelastic effects have been extensively studied in quantum dots and
nanotubes~\cite{Inoshita93,Inoshita97,Liang,Nygard,Kogan,Paaske:NatPhys2006}.
Recently, vibrational side-bands have been predicted
and reported for molecules displaying Kondo peaks at zero
bias~\cite{Cornaglia2004,Paaske:PRL2005,Ralph:PRL2007,Cornaglia2007,Pascual:PRL}.
Inelastic processes in Kondo physics have been studied by Zarand, Borda and
coworkers~\cite{LaszloPRL2004,LaszloPRB2007} who have shown that the
usual Kondo theories can reveal the amount of elastic and inelastic
spin-flip in the electron-impurity scattering event. For the particular
case of magnetic IETS, Zitko and Pruschke~\cite{Zitko:NJP2010} have
applied Kondo theories to the study of the coexistence of a Kondo peak
and of  IETS steps in the STM conductance of Co atoms on CuN/Cu(100)
substrates. Unfortunately, the numerical renormalization group method
that was used is not very accurate in reproducing the sharp conductance
steps. However, these authors {gave} the first unified picture of Kondo
and magnetic IETS.

More recently, Hurley and collaborators~\cite{Hurley:ArXiv2011}
have used the Kondo Hamiltonian and perturbation theory
to analyze the magnetic IETS of Co and Fe on CuN/Cu(100).
The authors conclude that certain spike-like {structures at the inelastic thresholds in}
 experimental IETS are {actually due to a Kondo-like effect}.
However, spike-like structures { close to the inelastic thresholds can also be found}  due to
electronic heating {under high current conditions} as
recent experimental and theoretical reports~\cite{LothNatPhys2010,DelgadoPRB82_2010_134414,Gauyacq3} show.
Indeed, the conductance {changes}
 in a non-linear
way if a tunneling electron 
{probes} the adsorbate still
excited by {earlier}
tunneling electrons. 

Here, we study the magnetic {transitions between singlet and triplet configurations of a magnetic adsorbate.  
We use our previous study 
 on } adsorbed copper phthalocyanine~\cite{Korytar}
and parametrize it to study the role of different ingredients in
the spectral function. 
Our study considers two independent spins localized in two different
molecular orbitals whose interaction is simply given by a Heisenberg
term of exchange interaction $I$. This model contains similar physics
to the recent two-impurity Kondo system studied by Bork and
co-workers~\cite{Bork}.

In the first part of this work, we use our previous {one-electron}
theory~\cite{Lorente:PRL2009}
to evaluate the 
 magnetic IETS for a singlet-triplet excitation. In the
second part, we use the non-crossing approximation~\cite{Korytar} and the self-consistent
ladder approximation~\cite{Maekawa} 
{for the same model} system. 
 The sharp
behavior of the conductance steps due to IETS is accounted for by both
theories. However, many-body effects {appear to be very} important at the
inelastic thresholds. Depending on the parameters of the impurity, we 
{further}
find that the actual {excitation} thresholds can be substantially 
{shifted} when many-body effects
are included. This finding can have important consequences in
the determination of MAE and {more generally of adsorbate magnetic structures} 
from IETS.

\section{One-electron theory}

The magnetic excitation of magnetic impurities on a solid surface
using an STM was modelled as an electron-impurity collision in
Refs.~[\onlinecite{Lorente:PRL2009,Gauyacq2,Gauyacq3}].  During the
collision of the electron with the impurity, {the tunneling electron and adsorbate spins briefly couple together. A transient resonant state with a finite lifetime can be involved, but not necessarily; it could simply be a scattering state that fixes the tunneling symmetries. These symmetries  depend}
 on both the electron
energy and the electronic structure of the impurity. In the present case,
the electron energy is very small since typical magnetic excitations
are in the meV range. Hence, the impurity's electronic structure
at the Fermi energy is the relevant one during the collision. In
Refs.~[\onlinecite{Lorente:PRL2009,Gauyacq2,Gauyacq3}] the impurities
had orbitals straddling the Fermi energy of the substrate { which were thus involved in the 
tunneling process.}

In the present work, we extend the study to systems where a positive
ion is a more likely origin of the transient state. In particular, we
are studying the singlet-triplet excitations of 
{ a magnetic impurity}
with a large charging energy $U$. In this case, the negative
ion is energetically less accessible than the positive one. This
{feature} has been used to {model} the Kondo effect in certain molecular
systems~\cite{RouraBas,Roura:JPCM,CornagliaEPL2011,Korytar}.

The modelling of the Kondo effect runs parallel to the 
{one-electron} theory
used to account for magnetic excitations in impurities. Indeed, the
Kondo effect (see for example Ref.~[\onlinecite{Korytar}]  and references therein)
can be described as a fluctuation between a charged transient
state and the ground state of the impurity. 
{The Kondo effect builds up on the coherence between
impinging electrons, it is a genuine many-body effect, while our earlier
IETS studies~\cite{Lorente:PRL2009,Gauyacq2,Gauyacq3} only consider single-electron collisions.}

Both in the one-electron and many-body pictures, the {tunneling} state
connects the initial state of the impurity with a final state that
can be different, leading to magnetic excitations. We {study below
how many-body effects influence magnetic excitation processes.}

\subsection{Inelastic transition rate}

{We assume the tunneling process (electron-adsorbate collision) to be very fast, much faster than the interaction
 at play in the singlet-triplet splitting, so that we can resort to the sudden approximation. The}
$T$-matrix~\cite{Gauyacq2,Gauyacq3} between an initial state $|i\rangle$
and a final state $|f\rangle$ { of the complete electron+adsorbate system is obtained from the 
sudden $T$-matrix  expressed in the basis set of the 
intermediate states $|S_T,M_T\rangle$ (spherical spin symmetry is kept during the brief collision). In the present case, since the system ground state is a singlet, the tunneling symmetry can only be $S_T$=1/2.}

In the large $U$ limit, the collision takes place between a hole
and the impurity. 
{ The ground state of the molecule is a singlet, then the initial state
is a singlet times (tensorial product) a hole
of the conduction band. Hence,}
\begin{equation}
|i\rangle = \hat{c}_{\sigma} |0,0\rangle
\label{initial}
\end{equation}
{ where $\hat{c}_{\sigma}$ destroys an electron 
with spin $\sigma$, and the impurity in the ground state
is given by its spin $|S,M\rangle$ which is a singlet 
$|0,0\rangle$ . Similarly, the final state is}
\begin{equation}
|f\rangle = \hat{c}_{\sigma'} |S,M\rangle
\label{final}
\end{equation}
{ where the impurity is left in one of the $|S,M\rangle$ states and
the hole is in a $\sigma'$ state. Below, we assume the STM tip to be
unpolarized, so that we sum over the contributions for $\sigma$ and
$\sigma'$ spins.}

{ The sudden $T$-matrix reduces to a projection operator  on all
the $|S_T,M_T\rangle$ states times a common transmission amplitude,
$B$~\cite{Lorente:PRL2009,Gauyacq2}. In tunneling, the
 transmission probability density
 is proportional to the density of states of
the sample, $\rho (\omega)$. This still holds in more complicated situations as
shown in Ref.~[\onlinecite{Wingreen1992}]. Then, the transmission probability density
is proportional to a constant coming from the transmission
amplitude, $|B|^2$, times the density of states.
   Hence, the elastic probability density for an electron energy $\omega$,
$T_e$, is given by }

 \begin{equation}
T_e (\omega)  = \rho (\omega) \sum_{\sigma}  | \sum_{M_T} | B  \langle S_T, M_T| \hat{c}_{\sigma} |0,0\rangle|^2|^2 = 
\frac{| B  |^2 \rho (\omega)}{2}.
\label{elastico}
\end{equation}
 The inelastic contribution, $T_i$, contains 
three possible orientations, $M_f$,
 of the impurity's spin because it is a triplet.
Hence,
 
\begin{eqnarray}
T_i (\omega) &=& \rho (\omega) 
\sum_{\sigma, \sigma',M_f} |\sum_{M_T}  B 
\langle 0,0 | \hat{c}^\dagger_{\sigma} |S_T, M_T \rangle
\nonumber \\
&\times &
\langle S_T, M_T  | \hat{c}_{\sigma'} |1, M_f \rangle |^2
\nonumber \\
&=&\frac{3| B |^2 \rho (\omega)}{2}.
\label{inelastico}
\end{eqnarray}

At this point, we can stress that the inelastic conductance is associated
to both spin-flip ($\sigma$ $\neq$ $\sigma'$) and non spin-flip ($\sigma$
$=$ $\sigma'$) processes for the tunneling electron; excitation
of the triplet state by a tunneling electron without a change of the
electron spin is 50 \% less probable than excitation of a triplet
adsorbate state with a spin-flip of the tunneling electron.

From Eqs.~(\ref{elastico}) and (\ref{inelastico})
the relative contribution of the elastic part to the transmision probability density at an
inelastic threshold  is just
\begin{equation}
\frac{T_e}{T_e+T_i} = \frac{1}{4}
\end{equation}
and 3/4 is the relative
contribution of the inelastic part of the transmission probability density. 


In the spirit of the generalization 
of the Landauer transmission formula to include inelastic
transitions~\cite{Wingreen1992,Ness2006,Monturet2008}, it is possible to link~\cite{Wingreen1992} the transmission probability density to the projected density of states
on the magnetic adsorbate or spectral function, $A$, and we use, below, this link to compare the results of the one-electron and many-body studies. 
In the present one-electron approach, the spectral function, $A$, is obtained as:
\begin{equation}
A = \rho (\omega) \{\frac{1}{4} +\frac{3}{4} [f(\omega + I) + f(I-\omega)] \},
\label{Atotal}
\end{equation}
where $I$ is the singlet-triplet excitation energy and $f(\omega)$ is the Fermi function of the substrate.
 
\begin{figure}
\centering
\includegraphics[width=0.4\textwidth]{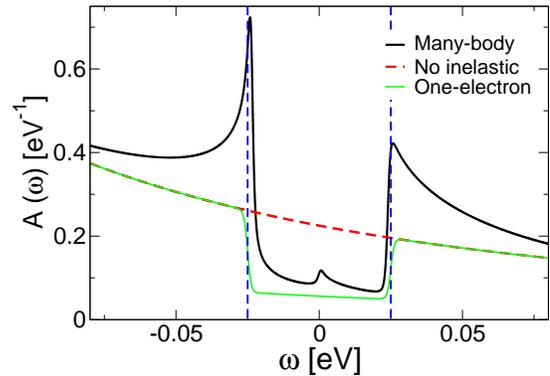}
\caption{\label{figure1}
Projected density of states on the magnetic adsorbate electronic structure
or spectral function $A$ as a function of the electron energy $\omega$
with respect to the Fermi energy of the substrate. The vertical dashed
lines show the one-electron inelastic thresholds. The figure presents
the adsorbate density of states in the absence of inelastic effects (marked
as ``No inelastic''),
the many-body spectral function and the one-electron conductance with
inelastic effects. The system's temperature is 7 K.} 
\end{figure}

Fig.~\ref{figure1} shows the characteristic step-like function for the
 spectral function, $A$, at low temperature. The two steps at positive and negative energy are associated to the opening of the inelastic triplet channel. The results in Fig.~\ref{figure1} use the parameterization 
of the adsorbed CuPc molecule determined in Ref.~\onlinecite{Korytar}, in particular the excitation energy, $I$, is equal to 25 meV.

\subsection{Thermal effects}

{The above one-electron results in Fig.~\ref{figure1}  correspond to
a low temperature of the surface. Thermal effects tend to wash out the inelastic
 structures via  two contributions}: i) the Fermi function is not
a step function and this rounds the conductance steps at threshold and
ii) at equilibrium at finite temperature, the system is not initially
entirely in the singlet state, a small thermal population of the triplet
is also present.

For the finite-temperature differential conductance,
the first effect has been discussed in detail in
Refs.~\cite{Lauhon-Ho2001,Lambe-Jaklevic}. The abrupt steps at
inelastic thresholds  that are present in the conductance at vanishing
temperature   are replaced by rounded steps due to the smearing of the
energy distribution of the electrons at finite temperature. 
This leads to a significant broadening of the step function of the order
of 5.5 $k_{B}T$~\cite{Lauhon-Ho2001}. In the present one-electron results
however, for a consistent  
comparison with the many body results on the spectral function, we
evaluate the above one-electron spectral function, i.e. we do not include the 
broadening effect coming from the thermal distribution of the electrons in the tip.



{The second effect is due to the finite thermal population of the triplet state by the
Boltzman factor, $F = \exp [-I/k_B T]$ and the corresponding decrease of the singlet population. The 
total spectral function, $A$,  is then equal to:

\begin{eqnarray}
A &=&  \{\frac{1}{4} +\frac{3}{4} [f(\omega + I) + f(I-\omega)] \} \frac{\rho (\omega)}{1+3F} \nonumber \\
& + &  \frac{3F \rho (\omega)}{1+3F}
\label{N_T}
\end{eqnarray}
}
Note the factor 3 due to the triplet
degeneracy. The total (elastic + inelastic) contributions of the singlet
and triplet states are equal. Since the triplet contribution  does not exhibit
any step,  the thermal population of the triplet state tends
to smooth out the stepped structure.  This effect is only visible on the results in 
Fig.~\ref{figure2} at the highest temperature.

\section{Many-body theory}

The extension of the above one-electron theory to the
many-body case {is} 
achieved by keeping the coherence between
impinging holes. This is a tremendous task, but can be easily
achieved using self-consistent schemes such
as the non-crossing approximation (NCA) and the self-consistent
ladder approximation (SCLA).

\subsection{Non-crossing approximation: Anderson Hamiltonian}

We consider that the impurity is fluctuating between two charged states: one
corresponding to the one-body ground state, and one of the above
transient states. The source of fluctuation is the hybridization
of the impurity with the substrate, $V$. By allowing the two charge
states to evolve self-consistently, NCA is an all-orders theory, albeit
neglecting all terms that ``cross''. It is at the sixth order in $V$
that the first crossing terms of the perturbation expansion are neglected.
NCA is then a method for the solution of the Anderson Hamiltonian,
where spin fluctuations are brought about by the hybridization term, $V$.

The considered Anderson-like Hamiltonian contains three terms,
\begin{subequations}
\label{Eq:model}
\begin{equation}
\label{And}
\hat H = \hat H_s + \hat V + \hat h.
\end{equation}

The first term is the free-electron-like Hamiltonian of the
substrate,
\begin{equation}
\hat H_s = \sum_\ksa\epsilon_{\mathbf k}
c^\dagger_\ksa c^{\phantom\dagger}_\ksa.
\end{equation}
The substrate electronic degrees of freedom are spin
$\sigma = \pm 1$, channel $a = 1,2$ and the remaining
degrees are encapsulated in the $\mathbf k$ symbol.

The impurity Hamiltonian
\begin{align}
\label{Eq:himp}
\hat h = \epsilon\sum_{a=1,2}\sum_\sigma
\ket{a\sigma}\!\!\bra{a\sigma}
+ I\mathbf S_1\cdot\mathbf S_2
\end{align}
has been used to describe a metal-organic adsorbate~\cite{Korytar}. 
The doubly degenerate ligand orbitals have orbital index $a=1,2$
and are represented by Hubbard operators, which project out
configurations where ligand occupation is higher than one.
The ligands are subject to exchange interaction with a third
orbital strongly localized at the molecular center which
is represented by a spin-half operator $\mathbf S_2$. Its charge
does not fluctuate because it is a very compact orbital
decoupled from the metallic substrate~\cite{Korytar}. The ligand spin operator can be expressed through
the vector of Pauli matrices $\bm\tau$ as follows
\begin{equation*}
\mathbf S_1 = \sum_{\sigma\sigma'}
\left(\frac{{\bm\tau}_{\sigma'\sigma}}{2}\right)\sum_{a=1,2}
\ket{a\sigma'}\!\!
\bra{a\sigma}.
\end{equation*}

The substrate - impurity hybridization is expressed by the term
\begin{equation}
\hat V = \sum_\ksa \left(V^{\phantom*}_{\mathbf k}
c^\dagger_\ksa \ket{0}\!\!\bra{a\sigma}
 + V_{\mathbf k}^*\ket{a\sigma}\!\!\bra{0}c^{\phantom\dagger}_\ksa\right)
\end{equation}
\end{subequations}
which does not mix spin, $\sigma$, and orbital ,$a$, degrees of freedom. The
empty ligand configuration is denoted by $\ket 0$.

Hence, Hamiltonian~(\ref{And}) describes~\cite{Korytar}: $(i)$ an electron
in an orbital disconnected from the reservoir, $(ii)$ the charge
fluctuations of the 2-fold degenerate orbital connected to the reservoir, and 
$(iii)$
the mutual interaction between both electrons
via an exchange term of matrix element $I>0$. This
system can have singlet-triplet excitations {with an excitation energy equal to I.}

The substrate-impurity hybridization enters in the NCA equations
via the energy-dependent width of the impurity orbitals. In
the present work, we  model it
by a rectangular function that is zero
for electron energies beyond the electron band, of bandwidth $2 D$.

For $I=0$ the spin $\mathbf S_2$ decouples  and we are left with
a SU(4) Kondo effect with a Kondo temperature $T_K^0$
given by~\cite{Korytar}:
\begin{equation}
k_B \; T_K^0\approx D e^{-\epsilon/ 4 \Gamma},
\label{temperature}
\end{equation}
where $2 \pi \Gamma$ is the level width of the orbital $a$ resonant
with the metal substrate.
 $T_K^0$ sets a natural energy scale of the present problem and
we will use it as the energy unit of the different calculated quantities.

\subsection{Self-consistent ladder approximation:
Coqblin-Schrieffer Hamiltonian}

It is also interesting to consider the above physical
model adapted to the Kondo Hamiltonian.
The Kondo Hamiltonian explicitly includes a spin-spin interaction
term between an itinerant electron and the impurity spin. In the
Kondo limit, $|\epsilon|\gg \Gamma$, where the impurity orbital energy, $|\epsilon|$, is much larger
than the level broadening, $\Gamma$ due to the hybridization term, $V$, 
both Anderson and Kondo Hamiltonian describe the same spin-flip physics~\cite{Wolff,Hewson}.
The Coqblin-Schrieffer Hamiltonian~\cite{Hewson,Coqblin} generalizes the
Kondo Hamiltonian to include orbital degrees of freedom, generalizing the
SU(2) Kondo problem to SU(N), where N is the combined orbital and spin
degrees of freedom of the impurity. The Coqblin-Schrieffer Hamiltonian is
interesting to be considered here because it allows us to explore
spin excitations in the Kondo limit. 
We will use the equivalent of NCA for the Coqblin-Schrieffer Hamiltonian,
namely the SCLA~\cite{Maekawa,Bickers:RMP}.

The Coqblin-Schrieffer Hamiltonian can be obtained from the
Anderson Hamiltonian, Eq.~(\ref{Eq:model}), in the Schrieffer-Wolff
limit
\begin{equation}
\label{Eq:CSLimit}
\epsilon \rightarrow -\infty,\quad \frac{\epsilon}{\Gamma} = const.
\end{equation}
Hence,
Hamiltonian \eqref{Eq:model} can be approximated by
the Coqblin-Schrieffer Hamiltonian~\cite{Coqblin}:
\begin{equation}
\hat H_{CS} = \hat H_s + \frac{V^2}{|\epsilon|^{\phantom 2}}
\sum_{\mathbf k\mathbf k'}
\sum_{aa'}\sum_{\sigma\sigma'}
\ket{a\sigma}\!\!\bra{a'\sigma'}
c^{        \dagger}_{\mathbf k',a'\sigma'}
c^{\phantom\dagger}_{\mathbf k,a\sigma}
 + I\mathbf S_1\cdot\mathbf S_2.
\label{CS}
\end{equation}

\subsection{Equilibrium regime}

The following results have been obtained using equilibrium NCA and SCLA. In principle, a non-equilibrium calculation is needed to account for the correct
coherence-decoherence balance at the excitation bias~\cite{RoschPRL2001}.
However, the situation studied here is the one found in STM studies
of molecules on surfaces. Typical parameters are tunneling currents below
the nA range and bias of a few meV. Assuming a single impurity resonance,
and a small bias, $V_{tip}$, such that the current takes place in resonance,
{the current can be estimated by a  Breit-Wigner-like  expression}

\begin{equation}
I \approx \frac{8 V_{tip} e^2}{h} \frac{\Gamma_{tip} \Gamma}{\Gamma + \Gamma_{tip}} .
\label{current2}
\end{equation}
where $\Gamma_{tip}$ is the resonance broadening due to hybridization
with the STM tip. 
If we take typical STM parameters such as 1 nA, 0.1 V, we obtain from Eq.~(\ref{current2}) that $\Gamma_{tip} \approx
10 \mu eV$. Typical $\Gamma$ for molecules on surfaces are larger than 100 meV
(see for example the calculations of Ref.~[\onlinecite{Korytar}]).
The non-equilibrium modification of NCA equations~\cite{HettlerPRB1998} for
slowly varying substrate density of states, reduces to the replacement
of Fermi occupation functions, $f (\omega - \mu)$, by an
effective distribution function $F_{eff}$ given by~\cite{HettlerPRB1998}:
\begin{equation}
F_{eff} (\omega) = \frac{\Gamma_{tip}}{\Gamma_{tip} + \Gamma} f (\omega - \mu_{tip})
+\frac{\Gamma}{\Gamma_{tip} + \Gamma} f (\omega - \mu).
\label{feff}
\end{equation}
Since $\Gamma_{tip}$ is easily a factor 1000 smaller than $\Gamma$ (a factor $10^4$
in the above example), we
recover $F_{eff}\approx f (\omega - \mu)$. Hence, in typical STM inelastic
measurements, the sample will be largely in equilibrium. {Qualitatively, the tip is extracting only an extremely small electron current from the molecule, much smaller than the electron fluxes that come from the substrate electron bath. If one further note that the bias applied to the junction is very small (tens of meV) compared to the local electrostatic potentials, then one can assume that the electronic structure of the probed molecule is not modified by the presence of the tip.} This is the
customary Tersoff-Hamann picture~\cite{Tersoff,LorentePRL2000}, where the
STM is assumed to read the unperturbed spectral function of the molecule on
the substrate.

\subsection{Singlet-triplet excitations}

Reference~[\onlinecite{Korytar}] gives a detail account for the implementation
of NCA in the case of singlet  and triplet molecules. 
As we showed in the previous section, the Hamiltonians contain an exchange term between
two spins localized in the molecule: 
\begin{equation}
\hat{H}_I = I {\bf S}_1 \cdot {\bf S}_2.
\label{HI}
\end{equation}
The two spins relate to
two different molecular orbitals that couple differently
with the metallic continuum. This gives rise to
a rich variety of physical situations depending on the
value of the exchange interation $I$, and the Kondo
scales of the different orbitals (see discussion in Ref.~[\onlinecite{Korytar}]
and references therein).

Here, we just consider positive
$I$, such that the molecular ground state is a singlet and
the singlet-triplet energy excitation is equal to $I$.  The case of $I$
$<$ 0  exhibits features similar to the ones described in the present
paper (see Ref.~[\onlinecite{Korytar}]).  Figure~\ref{figure1} shows
the results of the singlet molecule for an exchange interaction $I=25$
meV~\cite{Korytar}{; the Kondo temperature, $T_K^0$, is equal to 30 K}.  These results show that the general features of
the inelastic effect,{ i.e. the sharp conductance steps at inelastic
thresholds, } are readily understood with the above one-electron theory.

However, new features appear at the excitation thresholds that are purely of
many-body character.  Mainly, the spectral weight is greatly increased
near the thresholds.  This is reminiscent of the Kondo peak at zero
energy. {The difference is that in a SU(N) Kondo peak, there are fluctuations between degenerate impurity orbitals induced by  spin-flip  transitions at constant energy; in contrast, here, the singlet-triplet fluctuations involve an energy change of the adsorbate that is provided by the junction bias. In addition, in the present case, the singlet-triplet transitions are not pure spin-flip transitions, they can also occur without a change of the spin of the electron colliding on the impurity. }

The {structure} appearing at zero energy in the many-body spectral function is
due to the self-consistent approaches used here. The reason for its appearance
is the artificial flow of the marginal potential scattering term
in NCA~\cite{RoschPRL2001,KirchnerJLTP2002},
which overestimates potential scattering and hence leads to a spurious
Kondo-like feature at zero energy.

The results in Fig.~\ref{figure1} were obtained using the parametrization from
  Ref.~[\onlinecite{Korytar}] for the adsorbed CuPc. From now on in this paper, we will vary the parameters 
 in this modelling (ratio of excitation energy and Kondo temperature, energy of the orbital) in order to  
 decipher the role of the various parameters in the characteristics of the many-body
   features appearing close to the inelastic thresholds.

\subsection{Temperature effects}

\begin{figure}
\centering
\includegraphics[width=0.4\textwidth]{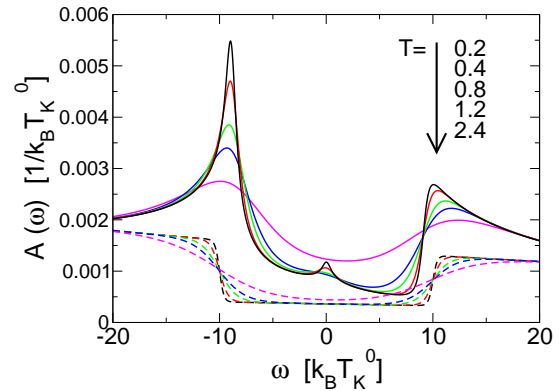}
\caption{\label{figure2} 
Temperature dependence of the spectral function for an
exchange coupling of the impurity's spins {equal to} ten times the Kondo
temperature, $I = 10\ {k_B} T_K^0$, for the Anderson model, Eq.~(\ref{And}). The energy axis is expressed in $T_K^0$ units, such
that the inelastic thresholds appear at $\pm 10$. 
As the temperature is raised (from 0.2 $T_K^0$ to 2.4 $T_K^0$), the Kondo-like
peaks {located close to the inelastic thresholds} are smeared and finally even the inelastic gap disappears. Dashed lines
are the results for the one-electron theory, that accounts
for the inelastic changes of the spectral function where all impurities
are assumed to initially be in their singlet state.
}
\end{figure}

When the temperature, $T$, is much smaller than the excitation energy
($k_B T \ll I$) thermal effects due to the 
{equilibrium} population of the triplet
state are negligible. The one-electron cases of Fig.~\ref{figure1} and
{to a lesser extent of Fig.~\ref{figure2}} are 
in this regime. Hence, the {dominant}  
{thermal effect is due
to the smearing of the sharp conductance steps at the inelastic
thresholds due to  the Fermi function broadening.}

Figure~\ref{figure2} is a systematic study of the temperature effect 
in the spectral
function with and without many-body 
effects for {$I=10\ {k_B} T_K^0$.}
As for Fig.~\ref{figure1},
the many-body results have been obtained for the Anderson Hamiltonian, Eq.~(\ref{And}) solved using NCA.
The many-body peaks appear to vary rapidly with T, they collapse as the temperature is increased.
This behavior, {for moderate temperatures (${k_B}T\ll I$)}, is due to the excitation
of thermal electron-hole pairs that destroy the electron coherence,
reducing the rate of coherent 
{singlet-triplet transitions} and hence the enhanced density
of states at the excitation thresholds, similarly to the disappearance
of zero-energy Kondo peaks with temperature.

The comparison of the one-electron and many-body spectral functions
shows that the {apparent energy thresholds are displaced one with respect to the}
other. This is already apparent in Fig.~\ref{figure1} where the threshold
coincides with the mid-point of the inelastic step for the one-electron
spectral function but {the mid-points of the inelastic steps in the many-body results are clearly shifted from the energy threshold}.  
Figure~\ref{figure2} shows a common point
where the one-electron spectral functions cross for the five considered
temperatures.  However, there is no common crossing point in
the many-body curves and their behavior is controlled by the temperature evolution of
the Kondo-like peaks.

\begin{figure}

\centering
\includegraphics[width=0.4\textwidth]{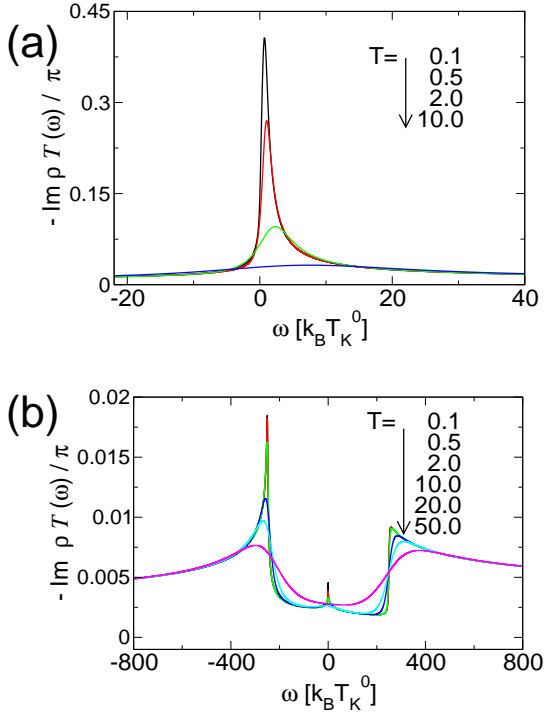}
\caption{\label{comparaisonI0}
Imaginary part of the $T$-matrix times the density of states $\rho$
for different temperatures in $T_K^0$ units.
$(a)$ SU(4) case for the spin 1/2 problem obtained when the
exchange interaction between the two localized spins is set to zero, $I=0$.
$(b)$ the singlet-triplet excitation for a
large value of the exchange integral, $I= 256\, k_B T_K^0$.
}
\end{figure}

For the SU(4) Kondo effect (case of a vanishing exchange interaction, $I$), the Kondo peak is greatly 
diminished at $T \approx T_K^0$.
This is seen in Fig.~\ref{comparaisonI0} $(a)$, where the $I=0$, SU(4) Kondo peak
is plotted for several temperatures. Figure~\ref{comparaisonI0} has been
obtained by solving the $T$-matrix, $\mathcal{T}$, for the Coqblin-Schrieffer Hamiltonian, Eq.~(\ref{CS}). 
The figure shows the imaginary part of the $T$-matrix times
the density of states $\rho$ as a function of the electron energy $\omega$.
This is equivalent to plotting the spectral function as a function
of $\omega$, since the hybridization function $\Gamma$ is the proportionality
factor connecting them.
Comparison of Figs.~\ref{figure2} and \ref{comparaisonI0} shows the equivalence
of both Hamiltonians, Eqs.~(\ref{And}) and (\ref{CS}), 
and of the solution methods. 

Surprisingly, when the temperature behavior of the inelastic Kondo-like
peaks is studied {(Fig.~\ref{comparaisonI0} $(b)$)}, we observe that the  peaks survive the temperature
increase much longer than the SU(4) peaks (Fig.~\ref{comparaisonI0} $(a)$).
Figure~\ref{comparaisonI0} $(b)$ shows 
$- Im \rho \mathcal{T}$ as in $(a)$, for $I=256\,k_BT_K^0$. However, we see that the spectral features are
still important at $T \approx T_K^0$, and even at $T = 50 \; T_K^0$ spectral peaks
are still visible.
However, at $T = 2.0 \; T_K^0$
the SU(4) peaks are very diminished, {Fig.~\ref{comparaisonI0}$(a)$}. This shows that
 {the inelastic process with  an  excitation energy equal to $I$ sets } 
in a new energy scale that {controls the spectral variation}
 with temperature. Only when $T_K^0$ is very small {(results not shown here)},
we recover the one-electron limit. Hence, the conditions where
many-body effects are not observable, {while inelastic effects are observable}  are
$k_b \; T_K^0 \ll k_b \; T \ll I$.

\begin{figure}
\centering
\includegraphics[width=0.4\textwidth]{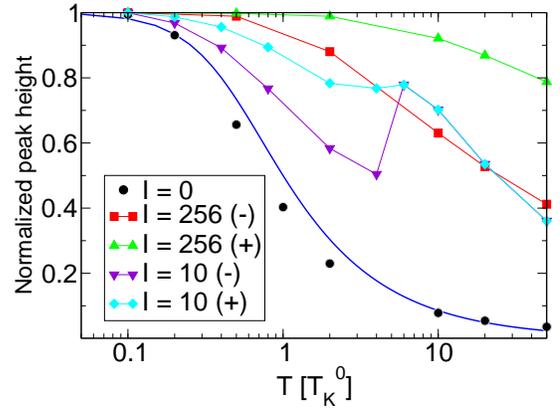}
\caption{\label{PeakHeights}
Maximum value of the spectral functions of Figs.~\ref{figure2} and
\ref{comparaisonI0} as a function of
temperature normalized to the value at $T\approx 0$. {  
Black dots: peak near the Fermi energy
for $I=0$  (Fig.~\ref{comparaisonI0}$(a)$); mauve down triangles: peak
near $\omega= -10\, k_B T_K^0$ and cyan diamonds: peak near $\omega= -10\, k_B T_K^0$ 
for  $I=10 \, k_B T_K^0$  (Fig.~\ref{figure2}); red squares: peak near
$\omega= -256 \, k_B T_K^0$  and green triangles: peak near $\omega= 256 \, k_B T_K^0$ for $I=256 \, k_B T_K^0$ (Fig.~\ref{comparaisonI0}$(b)$).}
The blue line is an empirical fit {to the results for $I = 0$ }
that displays at $T=T_K^0$ a maximum value
half {of that at  $T=0$.}
{The normalization of the down-triangle curve has been changed above $T= 6\, T_K^0$, due to the coalescence of the two inelastic Kondo-like peaks at high $T$ (see text for details).}
}
\end{figure}

For temperatures above $I$, the two inelastic peaks coalesce giving
rise to a {a single broad peak, similarly to the SU(4) limit. This 
variation appears in more detail in Fig.~\ref{PeakHeights} which presents the heights
 of the inelastic Kondo peaks relative to their value at low T as functions of the temperature.}

For $I=0$, the peak heights are approximately described by the blue
curve {(Fig.~\ref{PeakHeights})} given by:
\begin{equation}
A_{max}=\frac{1}{1+(2^{1/s}-1)(T/T_K^0)^s},
\label{fit}
\end{equation}
with $s = 0.4$. The same expresion, for $s=0.21$ has been used to fit the results
for the conductance in the Kondo regime from a renormalization group calculation~\cite{Goldhaber,Costi}.
Since, the fitted quantities ($T$-matrix {\em vs} conductance) and the
system (SU(4) {\em vs} SU (2)) are not the same, it is not surprising that $s$ is different.
This fit ensures that the peak has dropped to 1/2 at $T=T_K^0$,
which shows that the expression for the Kondo temperature, $T_K^0$ given by
Eq.~(\ref{temperature}) is a satisfactory approximation for both the NCA and SCLA. 

Figure~\ref{PeakHeights} also plots the {temperature evolution of the 
heights of the two inelastic Kondo peaks} 
for $I = 10\, k_B T_K^0$ ({down triangles and diamonds for lower and 
uppper peaks, resp.}) and $I= 256\, k_B T_K^0$ ({squares and up triangles 
for lower and 
uppper peaks, resp.). Each peak height is normalized to its value for $T\approx 0$}. At {$T\approx 6\, T_K^0$,} the two peaks of the $I = 10\, k_B T_K^0$ case 
coalesce and form a single
 peak. In order to show this clearly in Fig.~\ref{PeakHeights} we have to change the normalization
  of the peak heights above  6 $T_K^0$. We chose to normalize the single peak above  6 $T_K^0$ 
   as the high energy peak. This increases the
  discontinuity in the down-triangle curve but preserves the diamond curve. 
Beyond {$T\approx10\, T_K^0$} the curve
evolves more softly. This behavior shows that $I$ is the energy scale
that governs the spectral-feature evolution
at high temperatures. For the $I= 256\, k_B T_K^0$ cases, the plotted temperature range (see e.g. in Fig.~\ref{comparaisonI0} $(b)$) does not  reach a point at which the two peaks overlap significantly.

The different sensitivity to temperature increase in the inelastic case as compared
to the  'usual' Kondo problem, can be attributed to the different role
of electron-hole pairs. Decoherence increases as a temperature {rise} induces more electron-hole pairs. For the usual 
Kondo case, a Fermi electron is 
scattered by  the impurity,
hence thermal electron-hole pairs are very efficient in causing the
electron decoherence.
However, in the inelastic case, an excited electron is 
scattered by the impurity and { so it concerns the energy range around I}.
Hence, decoherence becomes particularly efficient when the thermal electron-hole
pairs have enough energy to reach the inelastic transition range, i.e. when the temperature is of the
order of the excitation energy. This explains why as we increase $I$,
the Kondo-like peaks survive at higher temperatures.

\subsection{Orbital energy dependence of the excitation threhold}

The inelastic spectra show Kondo-like features at the excitation
thresholds. Zero-energy Kondo peaks are basically determined by the
value of the Kondo temperature, $T_K^0$, and of the orbital energy, $\epsilon$.
$T_K^0$ is responsible for the zero-temperature width of the Kondo peak
and $\epsilon$ is an important factor 
for the actual shape of the Kondo
peak by virtue of the Friedel-Langreth sum rule~\cite{LangrethPR1966,Hewson}.
For this reason, we have studied the variation of the inelastic features
with the orbital energy, keeping the Kondo temperature, $T_K^0$, constant.

\begin{figure}
\centering
\includegraphics[width=0.4\textwidth]{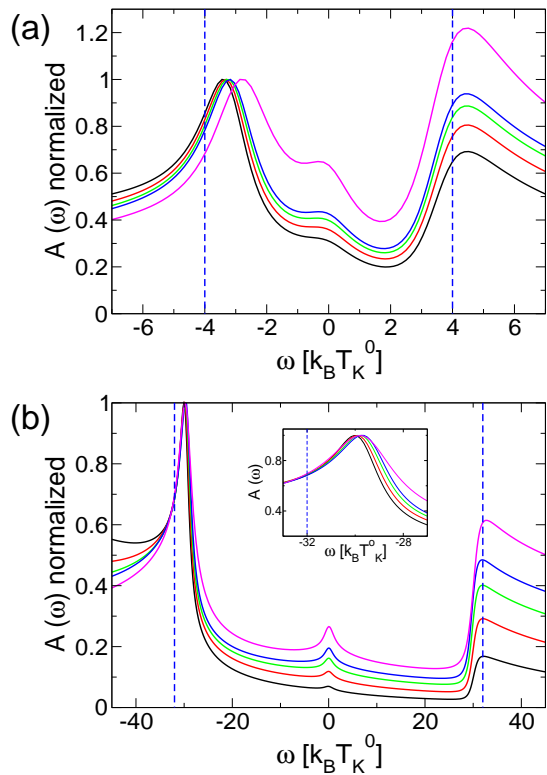}
\caption{
\label{figure3}
Normalized spectral function as a function of the electron
energy in units of the Kondo temperature, $T_K^0$. Case
(a) corresponds to an excitation energy $I=4\, k_B T_K^0$ as depicted
by the vertical dashed line. Case
(b) corresponds to the $I=32\, k_B T_K^0$ case.
The curves correspond to orbital energies $|\epsilon|= 110, 220,
440, 800 \,  k_B T_K^0$ for the curves in increasing spectral function in the
positive-energy part of the graphs. The topmost curve has been computed
using the equivalent Coqblin-Schrieffer Hamiltonian, and hence corresponds
to the $\epsilon \rightarrow -\infty$ case. The insets magnifies the
negative-energy peaks of the {$I=32\, k_B T_K^0$} case. The temperature is $T=0.1\, T_K^0$ } 
\end{figure}

Figure~\ref{figure3} shows the evolution of the inelastic features for
five different values of the orbital energy   
{(excitation energy $I$ equal to $4\, k_B T_K^0$ in Fig.~\ref{figure3}(a) and to $32\, k_B T_K^0$ in Fig.~\ref{figure3}(b)).}
 The first
four orbital-energy values increase by a factor of two: $|\epsilon|=
110,\, 220,\, 440$\, and $880 \; k_B T_K^0$. The fifth value is computed using the
Coqblin-Schrieffer Hamiltonian, Eq.~(\ref{CS}), that corresponds to the Kondo-limit or
$\epsilon \rightarrow -\infty$.  The negative-energy peaks of the spectral
functions have been normalized to one by multiplying the full spectral
function by a constant number.
The
behavior of the low and high- energy tails can be understood just by
the change in the density of states as the orbital energy shifts,
because
$T_K^0$ is kept constant
and hence,
the ratio $\epsilon/\Gamma$ is constant, where $2 \pi \Gamma$ is the width
of the one-electron peak originating at $\epsilon$. Hence, as $\epsilon$
approaches $-\infty$, the width $2 \pi \Gamma$ increases. For
$|\epsilon|=
110 \; k_B T_K^0$, the orbital resonance is close to the Fermi energy,
and hence, the density of states drops rapidly, while for a larger
value, the resonance is far from the Fermi energy and much broader,
dropping more slowly following the trends of  Fig.~\ref{figure3}.

Despite the relation of the peaks at threshold with Kondo features,
it is difficult to conclude on some type of extension of the Fermi-Langreth
sum rule since the two peaks at the inelastic thresholds 
present different behaviors.

As $|\epsilon|$ increases the peaks at thresholds move away from
the threshold energy towards positive energies. However, the inset
shows that the $\epsilon \rightarrow -\infty$ case for $I=32\; k_B\, T_K^0$
is anomalous in the sense that it presents an extra broadening instead
of a positive-energy shift. This hints at some saturation effect as the
energy scales increase.

It is noteworthy that the evolution of the negative-energy peaks is
faster than the positive ones.  The peaks at negative energy are also
narrower than the positive energy ones and their relative heights change
depending on the values of $I$ and $\epsilon$. While for $I= 4 \; k_B T_K^0$
the positive-energy peak is equal or larger than the negative-energy
case, for $I= 32 \; k_B T_K^0$ they are smaller. 

Figure~\ref{figure3} also shows that the deviation of the
peaks from the inelastic thresholds, $\Delta I$, depends on $I$. 
 We study this behavior in the next section.

\subsection{Asymptotic behavior of the threshold renormalization}

All the above results show that the {apparent inelastic }thresholds shift as $I$ increases. Actually,
the threshold shifts lead to a reduction of the inelastic gap. We can
quantify this reduction by studying the appeareance of singularities
in the resolvents that translate into peaks of the spectral function. In
order to perform this study, we have used the above SCLA
applied to the Coqblin-Schrieffer model.

Let $E_0$ and $E_1$ be the bare energies of the spin zero
and spin one multiplets. 
 The dressed multiplet energies $E_0^*,E_1^*$
are given by solving the equations
\begin{align*}
E_0^* &= E_0 + \Re \left\{\Sigma (E_0^*)\right\}\\
E_1^* &= E_1 + \Re \left\{\Sigma (E_1^*)\right\},
\end{align*}
where $\Sigma(\omega)$ is the pseudo-fermion self-energy.
Since $I$ is equal to the singlet-triplet
excitation energy, $I = E_1-E_0$, the threshold shift $\Delta I$
is given by
\begin{align}
\Delta I = (E_1^*-E_0^*) - (E_1 - E_0) = \Re\left\{\Sigma (E_1^*)-\Sigma (E_0^*)\right\}
\end{align}

\begin{figure}
\centering
\includegraphics[width=0.4\textwidth]{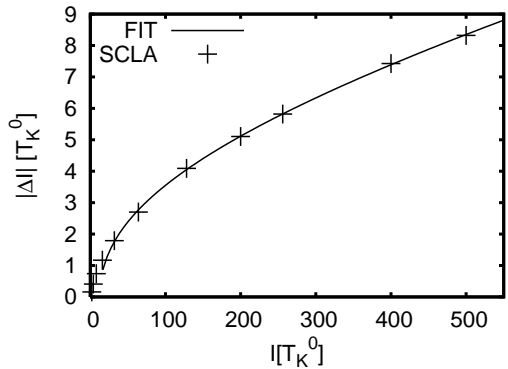}
\caption{
Shift of the excitation energy, $|\Delta I|$, versus excitation energy
$I$, crosses. All results are in $T_K^0$ units. Full line, fit {of the numerical results} to
the function $|\Delta I| = \sqrt{a x^2 + b x + c}$ where $x = I / ln
(I/k_B\,T_K^0)$. Hence, for values of $I$ larger than the ones of the present graph,
the threshold shift, $|\Delta I|$, follows the asymptotic behavior $\sim I
/ ln (I /k_B\,T_K^0)$.  } 
\label{figure4}
\end{figure} 

Figure~\ref{figure4} 
 shows the absolute value of the threshold shift as
a function of the excitation energy $I$ in $T_K^0$ units. The function
$|\Delta I| = \sqrt{a x^2 + b x + c}$
with $x = I / ln (I)$ is an excellent fit for a large range of values of $I$.
The fitting parameters are  $a= 0.0029$, $b= 0.6700$ and $c= - 3.3361$.
Hence, for asymptotically large excitation energies, $|\Delta I|$
is of the order of {$\sqrt{a}\ I / ln (I/k_B\,T_K^0)$. }This result is completely equivalent
to the asymptotic behavior found for the shift of Kondo peaks under
a magnetic field $B$ as estimated by Rosch and co-workers~\cite{RoschPRB2003}
and by studying the shift of spinon density of states under a
magnetic field by Moore and Wen~\cite{MoorePRL2000}.

\section{Discussion and conclusions}

The above results show that {many-body effects in }magnetic IETS can be readily
studied with usual Kondo-physics tools such as the self-consistent
approaches NCA (for the Anderson model) and SCLA (for the Coqblin-Schrieffer
one). Comparison with one-electron approaches~\cite{Lorente:PRL2009} is
in overall agreement and permits us to discern many-body effects
that appear as a consequence of  
{the impurity excitation's} coherence when the
inelastic channels open.  Here, we have studied the case of
a singlet-triplet excitation, when the tunneling electrons have
enough energy to overcome the singlet-triplet energy difference, $I$.
This situation has been achieved experimentally
 in carbon nanotubes~\cite{Paaske:NatPhys2006}. Related experiments
are the ones performed  in Mn dimers~\cite{HeinrichScience2006}. 
However, we expect these results to be {of relevance for} 
 many IETS cases
when the energy scales correspond to the Kondo ones.

The one-electron magnetic IETS theory runs parallel to the NCA. Namely,
a spin excitation takes place {via} a charge transfer process. During
this process the magnetic impurity changes its charge state. The
decay of the charge state back into the adsorbed state can lead
to a final state different from the initial one. This is the essence
of the excitation process. NCA builds on the same idea in a self-consistent
way such that charge exchange processes are included to all orders,
keeping their coherence. Hence, Kondo-like features are
included. 

The many-body features revealed in this study can be summarized {by} 
 the distorsion of the spectral properties of the impurity as compared to the
one-electron case. 
{Two peaks appear close to} the inelastic thresholds that are due to 
spin fluctuations when
at least two spin states become degenerate { similarly to the 'usual'  Kondo case}.
The inelatic peaks are shifted with respect
to the energy degeneracy point (the inelastic threshold). This leads
to a narrowing of the inelastic gap.
 This narrowing increases as $I$ increases following
the asymptotic law {$|\Delta I |\sim I/ln(I/k_B\ T_K^0)$.} This is exactly the
behavior found for a Kondo peak split {by 
a magnetic} field~\cite{MoorePRL2000}.

The resemblance of the present results with {those for the Kondo effect 
in the presence of magnetic fields~\cite{DickensJPCM2001,RoschPRB2003,MoorePRL2000} is due to 
the similarities between the two physical processes: }in both cases, there is a magnetic
excitation, in the present case due to the interaction between
two localized spins, and in the magnetic-field case due to Zeeman
energy splitting, and when the electron energy is large
enough to open the excited channel, the ground and excited states
are connected via inelastic spin-flip electron collisions (note that in the present case involving singlet-triplet transitions, these are both  of spin-flip and non-spin-flip type).

Finally, the many-body effects described here should 
 be observable at large
enough Kondo temperatures, $T_K^0$. This implies that the
molecule should be in the Kondo regime ($|\epsilon| < \Gamma$) but
with a sizable  $T_K^0$ ({\it i.e.} $|\epsilon|/ \Gamma \sim O(1)$
). In this situation, peaks at the inelastic thresholds will
be of many-body nature, leading to a strong renormalization of
the thresholds (see Fig.~\ref{figure4}) which can have important consequences
in the use of magnetic IETS as a {spectroscopic tool}.
The actual observation of many-body features may however depend on the
measuring procedure. STM measurements involve several orbitals that may or
may not be involved in Kondo physics. For realistic systems, the identification
of spectral function with measured conductance is sometimes not
straightforward. Although the Kondo peak may prevail
in the spectral function, the multi-orbital character of
the STM conductance can lead to channel interference and other
effects with the consequent appearance of complex Fano profiles~\cite{Crommie,Zawadowski}.

Our results show that once  Kondo-like features are
present, they are more robust than usual Kondo peaks. Indeed, while
Kondo peaks {completely} disappear at temperatures a few times the Kondo temperature,
inelastic many-body features {survive in this temperature range, if the  temperature is smaller than the
excitation energy.}
{Actually}, two energy
scales determine the many-body properties of IETS: the coherence
scale given by {$k_B\ T_K^0$,} and the excitation energy. 
Hence, not surprisingly 
as noticed by Hurley and coworkers~\cite{Hurley:ArXiv2011}, some
experimental IETS show spike-like features at the IETS thresholds, where
the results of our present work should be taken into consideration.

\bibliography{references}
\end{document}